# Bump Morphology of the CMAGIC Diagram

L. Aldoroty[1], L. Wang[1], P. Hoeflich[2], J. Yang[1], N. Suntzeff[1], G. Aldering[3], P. Antilogus[4], C. Aragon[3,5], S. Bailey[3], C. Baltay[6], S. Bongard[4], K. Boone[3,7,8], C. Buton[9], Y. Copin[9], S. Dixon[3,7], D. Fouchez[10], E. Gangler[9,11], R. Gupta[3], B. Hayden[3,12], Mitchell Karmen[3], A. G. Kim[3], M. Kowalski[13,14], D. Küsters[7,14], P.-F. Léget[4], F. Mondon[11], J. Nordin[3,13], R. Pain[4], E. Pecontal[15], R. Pereira[9], S. Perlmutter[3,7], K. A. Ponder[7], D. Rabinowitz[6], M. Rigault[9], D. Rubin[3,16], K. Runge[3], C. Saunders[3,7,17,18], G. Smadja[9], N. Suzuki[3,19], C. Tao[10,20], R. C. Thomas[3,21], and M. Vincenzi[3,22]

[1] George P. and Cynthia Woods Mitchell Institute for Fundamental Physics and Astronomy, Department of Physics and Astronomy, Texas A&M University, College Station, TX, 77843, USA; laldoroty@tamu.edu
[2] Department of Physics, Florida State University, Tallahassee, Fl, 32306, USA
[3] Physics Division, Lawrence Berkeley National Laboratory, 1 Cyclotron Road, Berkeley, CA, 94720, USA
[4] Laboratoire de Physique Nucléaire et des Hautes Energies, CNRS/IN2P3, Sorbonne Université, Université de Paris, 4 place Jussieu, 75005 Paris, France
[5] College of Engineering, University of Washington 371 Loew Hall, Seattle, WA, 98195, USA
[6] Department of Physics, Yale University, New Haven, CT, 06250-8121, USA
[7] Department of Physics, University of California Berkeley, 366 LeConte Hall MC 7300, Berkeley, CA, 94720-7300, USA
[8] DIRAC Institute, Department of Astronomy, University of Washington, 3910 15th Avenue NE, Seattle, WA, 98195, USA
[9] Univ Lyon, Université Claude Bernard Lyon 1, CNRS/IN2P3, IP2I Lyon, F-69622, Villeurbanne, France
[10] Aix Marseille Univ, CNRS/IN2P3, CPPM, Marseille, France
[11] Université Clermont Auvergne, CNRS/IN2P3, Laboratoire de Physique de Clermont, F-63000 Clermont-Ferrand, France
[12] Space Telescope Science Institute, 3700 San Martin Drive Baltimore, MD, 21218, USA
[13] Institut für Physik, Humboldt-Universitat zu Berlin, Newtonstr. 15, D-12489 Berlin, Germany
[14] DESY, D-15735 Zeuthen, Germany
[15] Centre de Recherche Astronomique de Lyon, Université Lyon 1, 9 Avenue Charles André, 69561 Saint Genis Laval Cedex, France
[16] Department of Physics and Astronomy, University of Hawai'i, 2505 Correa Road, Honolulu, HI, 96822, USA
[17] Princeton University, Department of Astrophysics, 4 Ivy Lane, Princeton, NJ, 08544, USA
[18] Sorbonne Universités, Institut Lagrange de Paris (ILP), 98 bis Boulevard Arago, 75014 Paris, France
[19] Kavli Institute for the Physics and Mathematics of the Universe, The University of Tokyo Institutes for Advanced Study, The University of Tokyo, 5-1-5 Kashiwanoha, Kashiwa, Chiba 277-8583, Japan
[20] Tsinghua Center for Astrophysics, Tsinghua University, Beijing 100084, People's Republic of China
[21] Computational Cosmology Center, Computational Research Division, Lawrence Berkeley National Laboratory, 1 Cyclotron Road, Berkeley, CA, 94720, USA
[22] Institute of Cosmology and Gravitation, University of Portsmouth, Portsmouth, PO1 3FX, UK

## Abstract

We apply the color–magnitude intercept calibration method (CMAGIC) to the Nearby Supernova Factory SNe Ia spectrophotometric data set. The currently existing CMAGIC parameters are the slope and intercept of a straight line fit to the linear region in the color–magnitude diagram, which occurs over a span of approximately 30 days after maximum brightness. We define a new parameter, $\omega_{XY}$, the size of the "bump" feature near maximum brightness for arbitrary filters $X$ and $Y$. We find a significant correlation between the slope of the linear region, $\beta_{XY}$, in the CMAGIC diagram and $\omega_{XY}$. These results may be used to our advantage, as they are less affected by extinction than parameters defined as a function of time. Additionally, $\omega_{XY}$ is computed independently of templates. We find that current empirical templates are successful at reproducing the features described in this work, particularly SALT3, which correctly exhibits the negative correlation between slope and "bump" size seen in our data. In 1D simulations, we show that the correlation between the size of the "bump" feature and $\beta_{XY}$ can be understood as a result of chemical mixing due to large-scale Rayleigh–Taylor instabilities.

*Unified Astronomy Thesaurus concepts:* Supernovae (1668); Type Ia supernovae (1728); Photometry (1234); Spectrophotometry (1556)

## 1. Introduction

Type Ia supernovae (SNe Ia) are important to cosmology (Riess et al. 1998; Perlmutter et al. 1999) because they may be used to determine luminosity distances to their host galaxies due to the predictability of their light curves (Pskovskii 1967; Phillips 1993; Riess et al. 1996; Goldhaber et al. 2001). Due to this predictability, photometric data from SNe Ia are standardizable for cosmological studies.

Several successful methods have been developed to quantify SNe Ia light curves, including the decline rate $\Delta m_{15}$ and stretch parameters (Pskovskii 1967; Phillips 1993; Perlmutter et al. 1997; Guy et al. 2005, 2007; Burns et al. 2014; Kenworthy et al. 2021). Statistical methods have also been used, including functional principal component analysis (He et al. 2018). These models can be improved if additional information is considered (Wang et al. 2009; Foley & Kasen 2011; Rose et al. 2021).

Correcting for the effects of dust extinction and reddening is a key part of calibrating SNe Ia, as it has significant cosmological consequences. As light from the SN passes through its host galaxy dust, the dust interferes and selectively removes more blue than red light. A similar effect occurs when the light traverses the Milky Way. Although Galactic dust reddening is generally well-measured (Schlafly & Finkbeiner 2011), it is more difficult to quantify the effects of dust from other galaxies. Further, as more SNe Ia are discovered, their diversity becomes more apparent, and disentangling extragalactic dust reddening from intrinsic color variation becomes more important. Theoretically, Hoeflich et al. (2017) showed that the mass and metallicity of the progenitor white dwarf (WD) can affect the intrinsic color of SNe Ia by as much as 0.1 mag in $(B - V)$. Therefore, it is necessary to establish robust, reddening-free color parameters for SNe Ia.

The color evolution of SNe Ia has been observed to be similar across different events and thus has been used to estimate host galaxy extinction (Lira 1995; Phillips et al. 1999). Wang et al. (2003) introduced the color–magnitude intercept calibration (CMAGIC) method as a way to utilize data taken in the month after maximum brightness in order to standardize SNe Ia. About one week after maximum brightness, the color–magnitude diagram (hereafter "CMAGIC diagram") for normal-bright SNe Ia (i.e., neither subluminous nor overluminous) displays a remarkably linear relationship in the rest-frame $B$ magnitude versus $B - V$, $B - R$, and $B - I$ colors, which lasts for two to three weeks. The slope of this region, $\beta_{XY}$, is independent from other measurable quantities. We can use this property to calibrate SNe Ia accurately and independently of other methods, with sensitivity to different systematic sources of error. It is interesting to explore CMAGIC because in the future, we may be able to use CMAGIC to calibrate SNe Ia lacking data around maximum light. Further, it has been shown that CMAGIC curves may be useful in helping to break the degeneracy between intrinsic color and reddening (Hoeflich et al. 2017). Conley et al. (2006) shows that cosmological results from CMAGIC are consistent with the current picture of cosmology, i.e., an accelerating flat universe with a cosmological constant. Similarly, Wang et al. (2006) shows that CMAGIC methods have a Hubble residual rms deviation of approximately 0.14 mag, comparable to methods that use the maximum brightness $B_{\max}$.

Wang et al. (2003) notes two different morphologies found in the CMAGIC diagram—one with a luminosity excess around the time of maximum brightness (the "bump" feature), and one without. The authors also note a bifurcation in slope distribution, which they suggest may be indicative of two progenitor channels. Chen et al. (2021) also observed a varying slope in color curves, creating a proxy for the color–stretch parameter $s_{BV}$ (Burns et al. 2014). Conley et al. (2006) discuss the "bump" feature in more detail, stating that the probability of a "bump" occurring increases as $B$-band stretch increases; however, it is still possible to find SNe with the same stretch where one has a "bump" and the other does not. They find that SNe with stretch values of $s > 1.1$ have a bump, and none with $s < 0.8$ have one. Those with $1.0 < s < 1.1$ have a 50% probability of having a bump, and SNe with $0.8 < s < 1.0$ have an approximately 8% chance of having a bump. Wang et al. (2006) notes that the difference between $B_{\max}$ and the CMAGIC parameter $B_{BV}$ is directly tied to the existence of a bump, and therefore may be an important consideration for color corrections. The CMAGIC method has also been applied to derive distances and dust reddenings of some well-observed SNe Ia (Wang et al. 2020; Yang et al. 2020).

Hoeflich et al. (2017) note that CMAGIC is useful for studying the intrinsic physical properties of SNe Ia because the locations of its distinguishing features are affected by the central density of the progenitors and the explosion scenario, and propose that variations in the slope may also point toward underlying SN physics. If the shape of the CMAGIC diagram points toward physics, and some SNe show a "bump" feature where others do not, it is important to quantify this shape variation because it may enhance our understanding of the intrinsic colors of SNe Ia.

In this paper we present empirical relations as well as theoretical results of the CMAGIC diagram, centered around the "bump" feature. In Section 2.1, we describe the Nearby Supernova Factory (SNfactory; Aldering et al. 2002) data set used in this work. In Section 2.2, we describe the functional principal component analysis (fPCA) light-curve fitting based on the results of He et al. (2018). Section 2.3 describes the spectral analysis procedures. Section 2.4 describes the fitting procedures, as well as defining one useful "bump" parameter, $\omega_{XY}$. Results and discussion of the study are in Section 3. First, we discuss the "bump" morphology in Section 3.1, followed by theory based on the 1D Hoeflich et al. (2017) model 23, modified to include mixing. Section 3.3 contains CMAGIC diagrams of light-curve templates, including those from the fPCA method (He et al. 2018), SNooPy (Burns et al. 2011), and SALT3 (Kenworthy et al. 2021). We vary the templates' parameters in order to reproduce the morphology identified in the data. We show that all three sets of templates are successful at reproducing the "bump" (or lack thereof). Finally, the results are summarized in Section 4.

## 2. Method

### 2.1. Data

Spectra from SNfactory (Aldering et al. 2002) were used for this analysis. Details about the SNfactory data set and data reduction can be found in Saunders et al. (2018) and Aldering et al. (2020). After correcting the observed spectra to the rest-frame, synthetic photometry for each SN was made using $BVRI$ filters from Bessell & Murphy (2012), which were calibrated to the Vega system using `alpha_lyr_stis_010.fits` from the CALSPEC database (Bohlin et al. 2014) (see Appendix A). These filters were chosen for ease of comparison to Wang et al. (2003). The zero-point for SNfactory data is kept hidden, thus, all magnitudes in this work are the calculated magnitude plus a constant. Cuts were then applied to the data, requiring that observations exist before maximum light, and that a minimum of three observations exist in the linear region in all three types of CMAGIC diagram (see Section 2.4).

We do not explicitly remove peculiar SNe Ia. This work includes a total of 85 SNe, where there are 31 in the "bump" group, 34 in the "no bump" group, and 20 in the "ambiguous" group.

### 2.2. fPCA Fitting

Light curves are fit using fPCA, as described by He et al. (2018).[23] It is advantageous to use PCA methods to fit complex curves, such as light curves, because the result is a parameterization of the curve that is a linear combination of orthogonal PC functions. Therefore, it is straightforward to propagate the errors (See Appendix B). The fitted light curves

---

[23] These templates can be used via `snlcpy` (Aldoroty et al. 2022), located at https://github.com/laldoroty/snlcpy.

**Table 1**
Results for a Two-sample KS Test ($D_{n,m}$) and Independent Two-sample $t$-test for the Parameters Presented in This Work, Based on the "Bump" vs. "No Bump" Samples

| Parameter | $D_{n,m}$ | $p_{D_{n,m}}$ | $t$ | $p_t$ |
|---|---|---|---|---|
| $\Delta m_{15,B}$ | 0.544 | $7.40 \times 10^{-5}$ | $-3.74$ | $3.89 \times 10^{-4}$ |
| $\Delta m_{15,V}/\Delta m_{15,B}$ | 0.64 | $9.56 \times 10^{-7}$ | 6.30 | $3.25 \times 10^{-8}$ |
| pEW Si II $\lambda$6355 | 0.55 | $4.18 \times 10^{-5}$ | $-3.01$ | $3.70 \times 10^{-3}$ |
| pEW Si II $\lambda$5972 | 0.49 | $4.07 \times 10^{-4}$ | $-4.77$ | $1.12 \times 10^{-5}$ |
| $\beta_{BV}$ | 0.75 | $2.40 \times 10^{-9}$ | $-6.63$ | $8.94 \times 10^{-9}$ |
| $\beta_{BR}$ | 0.43 | $2.96 \times 10^{-3}$ | $-3.62$ | $5.89 \times 10^{-4}$ |
| $\beta_{BI}$ | 0.44 | $2.60 \times 10^{-3}$ | $-3.45$ | $2.60 \times 10^{-3}$ |
| $\omega_{BV}$ | 0.94 | $1.22 \times 10^{-15}$ | 11.61 | $2.69 \times 10^{-17}$ |
| $\omega_{BR}$ | 0.62 | $2.03 \times 10^{-6}$ | 5.13 | $3.01 \times 10^{-6}$ |
| $\omega_{BI}$ | 0.57 | $2.31 \times 10^{-5}$ | 4.98 | $5.20 \times 10^{-5}$ |

**Note.** We correct for the "look-elsewhere effect" by dividing our significance level $\alpha = 0.05$ by the number of parameters in this table. Thus, our significance level is $\alpha_C = 0.005$. The first section shows the results for parameters independent of the bump. The second section shows the results for slope $\beta_{XY}$ of the linear region of the CMAGIC diagram, which we have shown to be strongly correlated with "bump" size (Figure 2). The third section shows results for "bump" size $\omega_{XY}$. The functions `ks_2samp()` and `ttest_ind()` from `scipy.stats` were used. The first column lists the tested parameter; the second and fourth columns show the test statistics; the third and fifth columns show the $p$-values for the test statistics in the columns to their left. For the Kolmogorov–Smirnov (KS) test, the null hypothesis is that the "bump" and "no bump" samples are drawn from the same distribution. No assumption is made about the distributions of the data. The two-sample $t$-test checks the null hypothesis that the mean value of the two groups is identical. This test assumes the data are normally distributed. We do not assume equal variance. We are able to reject the null hypothesis for all parameters.

are used to determine the location of the brightest point in the $B$ band, as well as the change in magnitude between peak $B$ band brightness and 15 days later, $\Delta m_{15,B}$. We use the fits from the light curves to compute the CMAGIC diagram for each SN and color combination. For this analysis, only the first two $B$- and $V$-band specific PC components from He et al. (2018) are used because these describe the majority of the variation in the light curves, and we found that including the third and fourth components resulted in unphysical fits for some SNe because the data were insufficient to constrain the fit realistically using this method.

### 2.3. Spectral Analysis

Pseudo-equivalent widths (pEWs), i.e., the depths of spectral features given a pseudo-continuum drawn around an individual feature, are calculated for all SNe using the data spectrum nearest to maximum brightness available in the $B$ band. Gaussian fits were applied to the $\lambda$6355 and $\lambda$5972 Si II lines using a bootstrapping method. Two regions, each 20 Å wide, were identified around either side of each absorption line, and endpoints were randomly drawn 225 times from these regions to determine the continuum for normalization. The final pEW is the area integrated under the Gaussian fit, and the error is the standard deviation of each set of area measurements. This method mirrors the procedure used by Galbany et al. (2015). pEW is used as a parameter for statistical tests in Table 1, and is shown in Figure 3.

### 2.4. CMAGIC

The CMAGIC diagram of an SN Ia shows its evolution in brightness as a function of color (Figure 1). After explosion, the SN grows brighter and bluer in optical wavelengths. At maximum brightness, it starts to redden linearly as it dims over the next ∼30 days, before turning around and becoming linearly bluer as it continues to dim. Some SNe Ia show a small luminosity excess around maximum brightness (Figure 1, left), where others do not (Figure 1, right). We refer to this luminosity excess as the "bump" feature. In this section, we discuss the methodology used to handle the two linear regions in the CMAGIC diagram, followed by quantifying the size of the "bump" feature.

#### 2.4.1. Linear Regions

Wang et al. (2003) found that there are two linear regions that occur shortly after maximum brightness in the CMAGIC diagram; the first begins 5–10 days after maximum, and ends at roughly 30 days. The second begins at around 40 days (shown in the left panel of Figure 1), although discussion of this region is outside the scope of this study. To fit the first linear region (hereafter "linear region") of the $B$ versus $B-V$, $B-R$, and $B-I$ CMAGIC diagrams for each SN, we used Levenberg–Marquardt least squares minimization via `mpfit` in Python (Moré 1978; Moré & Wright 1993; Markwardt 2009; Koposov 2017). The fits were performed such that $\chi^2$ was fixed to equal the number of degrees of freedom via scaling the errors, with different scalings for the two linear regions. The endpoints of the linear regions in the CMAGIC diagrams for all SNe were determined by visual inspection.

SNe with fewer than three observations in the linear region of any of the three diagrams ($B-V$, $B-R$, or $B-I$) were excluded, in order to allow for a minimum of one degree of freedom in all linear fits.

#### 2.4.2. Quantifying the Size of the "Bump" Feature

The size of the "bump" was quantified by identifying the $B-V$ color corresponding to $B$-band maximum brightness. Then, the CMAGIC diagram was normalized by the linear fit. The "bump" size is defined as

$$\omega_{BV} = (\beta_{BV}(B-V)_{\max} + B_{BV0}) - m_{B\max}, \quad (1)$$

where $m_{B\max}$ is the magnitude at maximum brightness in the $B$ band, $\beta_{BV}$ is the slope of the linear region from the fit (purple line in Figure 1), $(B-V)_{\max}$ is the color at the time of $B$-band maximum, and $B_{BV0}$ is the value of the fit line when $(B-V) = 0$. If there is a bump, $\omega_{BV}$ will be positive; if there is no bump, the value will be negative. Error propagation for $\omega_{BV}$ is described in Appendix C.

## 3. Results and Discussion

### 3.1. "Bump" Morphology

SNe are distinguishable in the CMAGIC diagram by the presence, or lack, of a luminosity excess relative to the linear region near $B_{\max}$, the maximum magnitude in the $B$ band. We have qualitatively divided our sample into three categories based on visual inspection: those with a bump, those without a bump, and those where it is ambiguous whether or not there is a bump. The last group includes those without enough data in this region to say definitively if there is a "bump" or not, and those that appear as if they might have a bump, but if they do, it is very small. This analysis includes a total of 85 SNe, where there are 31 in the "bump" group, 34 in the "no bump" group, and 20 in the "ambiguous" group.

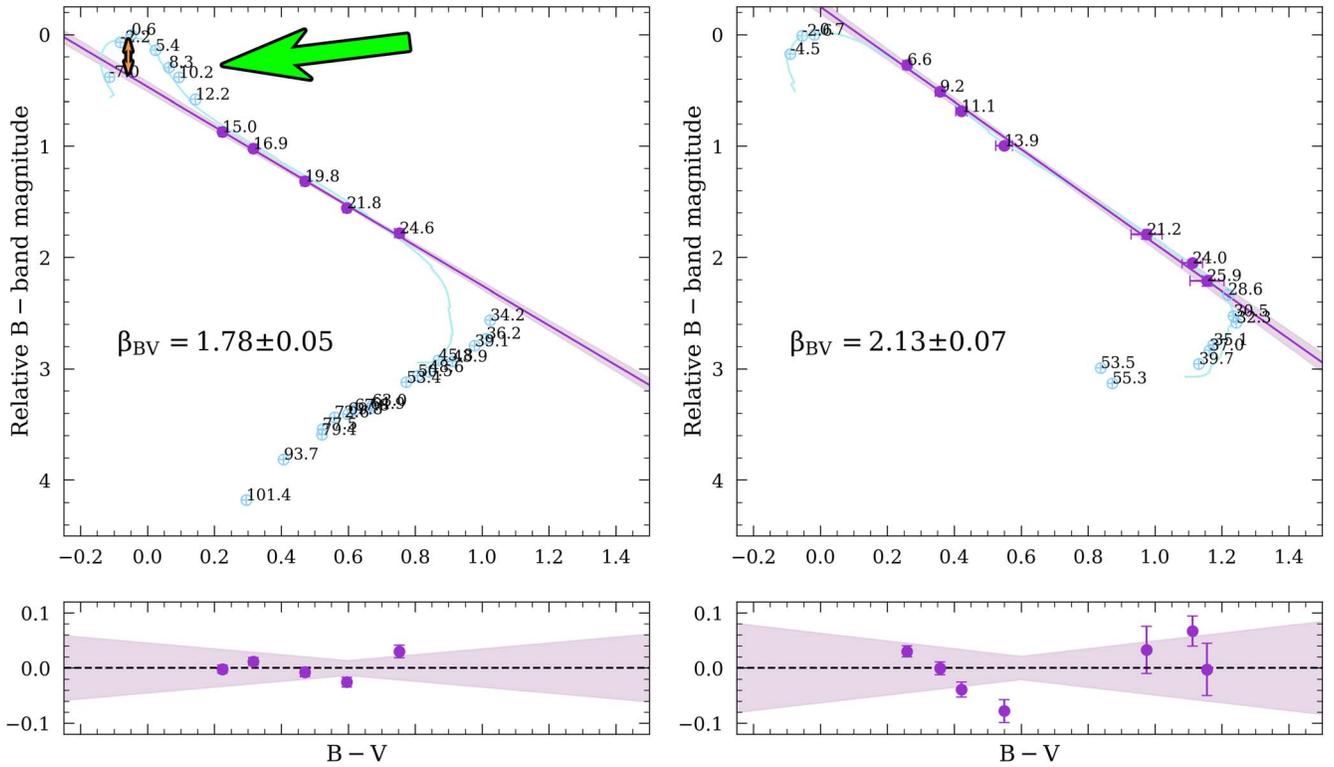

**Figure 1.** CMAGIC diagrams for two SNfactory SNe Ia. The left plot shows a "bump" feature. The green arrow points to the "bump" feature, and the small vertical orange arrow shows a visualization of the definition of $\omega_{BV}$ (Equation (1)). The right plot does not show a "bump" feature. The filled purple circles are members of the linear region and were therefore used in the linear fits. The open blue circles do not belong to the linear region, and were therefore excluded from the linear fits. The points are labeled by the light-curve phase relative to $B_{\max}$. The solid purple line is the linear fit and the solid blue curve is from the fPCA fit. $\beta_{BV}$ is the slope of the fit line (purple), and the fit was performed such that $\chi^2$ was fixed to equal the number of degrees of freedom via scaling the errors. Note that for the SN with the bump, data up to approximately 15 days after maximum brightness in the B band (the numbers tracking each data point) stay bluer than those for the SN without a bump.

The most striking difference between the two main categories is that SNe with a "bump" tend to have a smaller slope, $\beta_{XY}$, than those without (Figure 2) in all three of the CMAGIC diagrams analyzed here. A two-sample KS test run on the slopes of these two groups strongly suggests that they are likely to be drawn from different parent populations (Table 1). Further, the fact that the "ambiguous" category sits in the center of these indicates that the "bump" feature exists on a continuous basis rather than being discrete. We do observe an outlier: the isolated "no bump" data point in the first panel of Figure 2 is PTF10ygu (SN2010jn), a known bright SN Ia with a slow decline rate (Hachinger et al. 2013). Although it has more extreme values for $\omega_{BV}$ and $\beta_{BV}$ than the rest of the sample, this result is consistent with the rest of the trend. PTF10ops and SNF20070714-007 are the two "no bump" outliers in the second and third panels. PTF10ops has been shown to be subluminous and does not match expected explosion models (Maguire et al. 2011). The ``ambiguous'' outlier in the second and third panels is SN2004ef.

We find a separation between the "bump" and "no bump" categories when comparing the Si II $\lambda 6355$ and $\lambda 5972$ lines (Figure 3, Table 1), as well as $\Delta m_{15,B}$ (Figure 3). We confirm the tendency for SNe with a "bump" to be slower decliners, reflecting behavior that was first noted by Wang et al. (2006). If a SN has a bump, then its $B_{\max} - B_{BV}$ is larger (Wang et al. 2006), where $B_{\max}$ is the magnitude at maximum brightness in the B band and $B_{BV}$ is the B-band magnitude on the line fit to the linear region where $(B - V) = 0$. This quantity is strongly linked to intrinsic color. Wang et al. (2006) report a piecewise relation between $B_{\max} - B_{BV}$ and $\Delta m_{15,B}$ where $|B_{\max} - B_{BV}|$ decreases until it hits 0 when $\Delta m_{15,B} \approx 1.1$, where it then stays consistent with 0. In the right column and third row of Figure 3, we show a steep decrease in the number of SNe with a "bump" where $\Delta m_{15} \approx 1.1$, reflecting the behavior described by Wang et al. (2006).

We also note a strong separation between the "bump" and "no bump" categories in the ratio of $\Delta m_{15,V} / \Delta m_{15,B}$ (Figure 3, bottom right). Like the slope, a KS test indicates these are likely to be drawn from different samples (Figure 3). Once again, the fact that the "ambiguous" category lies in the center of these indicates a continuum of "bump" size rather than discrete types.

Our results may be compared to those of Chen et al. (2021), who report a similar color-related linear feature as found by Wang et al. (2003). However, their slope parameter $s_{0*}$ is derived from color as a function of time rather than the CMAGIC diagram. Chen et al. (2021) find a strong linear correlation between $s_{0*}$ and $s_{BV}$ (Burns et al. 2014), indicating that their $s_{0*}$ may be used as a proxy for $s_{BV}$. Our slope, $\beta_{XY}$, appears similar to their $s_{0*}$, however, we do not find any correlation between $\beta_{XY}$ and $s_{0*}$ nor $s_{BV}$, indicating that $\beta_{XY}$ contains independent information (Figure 4); i.e., that $s_{0*}$ is not able to discern the presence of a "bump" feature.

We compare the parameters defined in this work to the parameters from SALT3 (Kenworthy et al. 2021) in Figure 5. We find a weak negative correlation between $\beta_{BR,1}$ and $x_1$ in

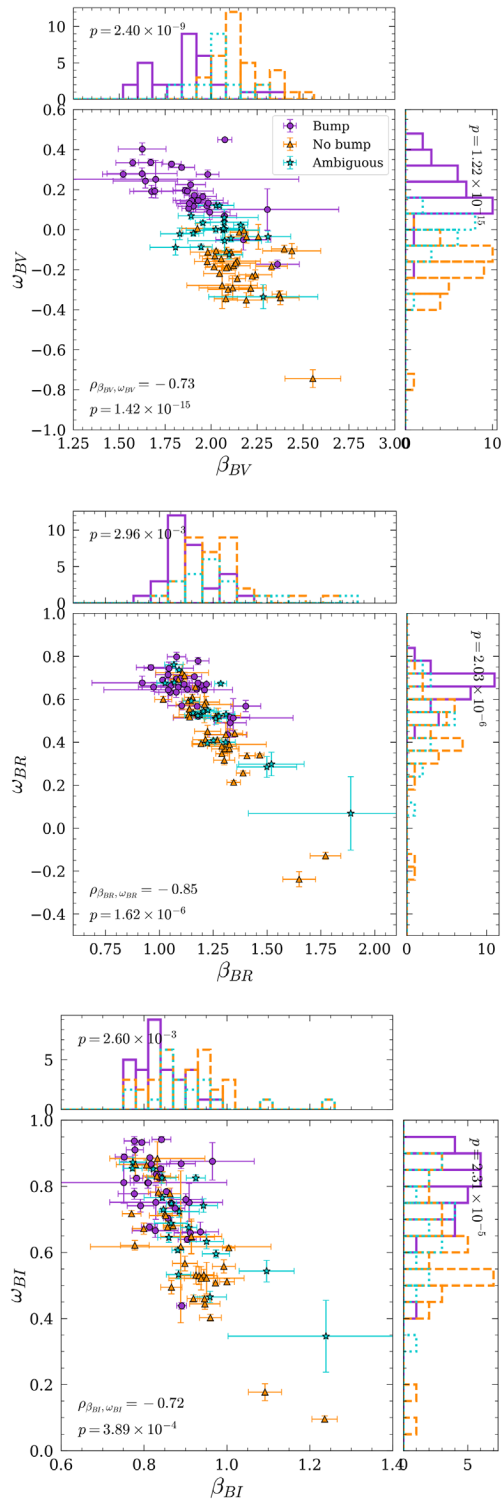

**Figure 2.** Correlation between "bump" size, $\omega_{XY}$, and slope, $\beta_{XY}$, for the $B-V$, $B-R$, and $B-I$ CMAGIC diagrams. SNe are always assigned to a "bump" or "no bump" category based on their $B-V$ CMAGIC diagram because every SN has a "bump" in the $B-R$ and $B-I$ diagrams. The purple circles represent the "bump" category, orange triangles represent the "no bump" category, and green stars represent the "ambiguous" category. The overlaid histograms show the frequency of the aforementioned categories, with the same color indications. The $p$-values in histograms represent the results of a two-sample KS test on the "bump" and "no bump" groups for the histogram it overlays. A smaller $p$-value means it is more likely that the samples are drawn from different distributions. $\rho_{X,Y}$ is the Pearson correlation coefficient, and the $p$-value below $\rho_{X,Y}$ is the corresponding $p$-value. Results from statistical tests for $\omega_{XY}$ and $\beta_{XY}$ are shown in Table 1.

Figure 5, corresponding with SNe having a slower brightness decline (i.e., brighter) evolving in color less drastically than fast decliners (i.e., slower). This reflects the tendency of SNe Ia with broader light curves to be brighter (Höflich et al. 1996). However, we caution that these relationships are not conclusive and will require further verification.

### 3.2. Theory

The CMAGIC diagrams of SNe Ia depend on the type of explosion and associated physics (Wang et al. 2003; Hoeflich et al. 2017). In this section, we demonstrate that inhomogeneous chemical mixing at the chemical interfaces can produce the observed variations in 1D models. We choose the delayed-detonation (DDT) scenario (Khokhlov 1989) because this class of models have been shown to reproduce the observations of the CSP sample (Hoeflich et al. 2017), and consider a typical model for a normal-bright SNe Ia. We do not fine-tune parameters such as the total WD mass, the burning properties, nor fit any individual objects. This is not necessary because it was shown by Hoeflich et al. (2017) that the templates agree with the observations, and that the brightness shift goes with the DDT transition density $\rho_{tr}$.

Rayleigh–Taylor instabilities during the deflagration provide a natural scale of about 1000–2000 km sec$^{-1}$ for chemical inhomogeneities. The subsequent detonation phase burns away the initial chemical inhomogeneities, except at the chemical interfaces, and spherical symmetry of the density is conserved (Gamezo et al. 2005; Wang et al. 2007; Cikota et al. 2019). The details of the flame propagation depend on the ignition condition in the WD, possibly magnetohydrodynamical effects (Khokhlov 1995; Niemeyer & Hillebrandt 1995; Remming & Khokhlov 2014; Hristov et al. 2018, 2021) and, in particular, the duration of the detonation phase. Initially, large-scale plumes are formed (Gamezo et al. 2003), which for prolonged deflagrations, decay to small scales (Röpke et al. 2007).

The "bump" in CMAGIC occurs around maximum light when the photosphere has receded to the Si–Ni interface for normal-bright SNe Ia (Höflich et al. 2002). Guided by detailed simulations and using the radiation hydrodynamical code HYDRA (Hoeflich 2002), we studied the effect on CMAGIC for (a) the unmixed model, (b) a large inhomogeneous scale mixing of the Si–Ni interface with a covering factor of about 50% (i.e., 50% of the material is rising and 50% of it is sinking) as representative of short deflagration phases, and (c) homogeneous mixing on the same scale.

A comparison between cases (a) and (b) is shown in Figure 6. For inhomogeneous mixing, in the transition region, plumes without mixing are transparent, and plumes with mixing are not. This creates a "picket fence"-like effect in the photosphere. Radiation can escape through these transparent regions, causing a small luminosity excess, i.e., the "bump" effect we observe in the data. Once the photosphere has receded past the mixed region, some energy has already escaped through the plumes, causing a slower rate of energy loss. A slower rate of energy loss corresponds to the shallower slope observed in the "bump" population.

Overall, the opacity increases for case (c). This results in redder colors throughout the CMAGIC curve; this is because at maximum, the photosphere has already receded well within the high opacity Ni/Co/Fe-rich region. Due to the lack of a "picket fence" effect, no "bump" is produced.

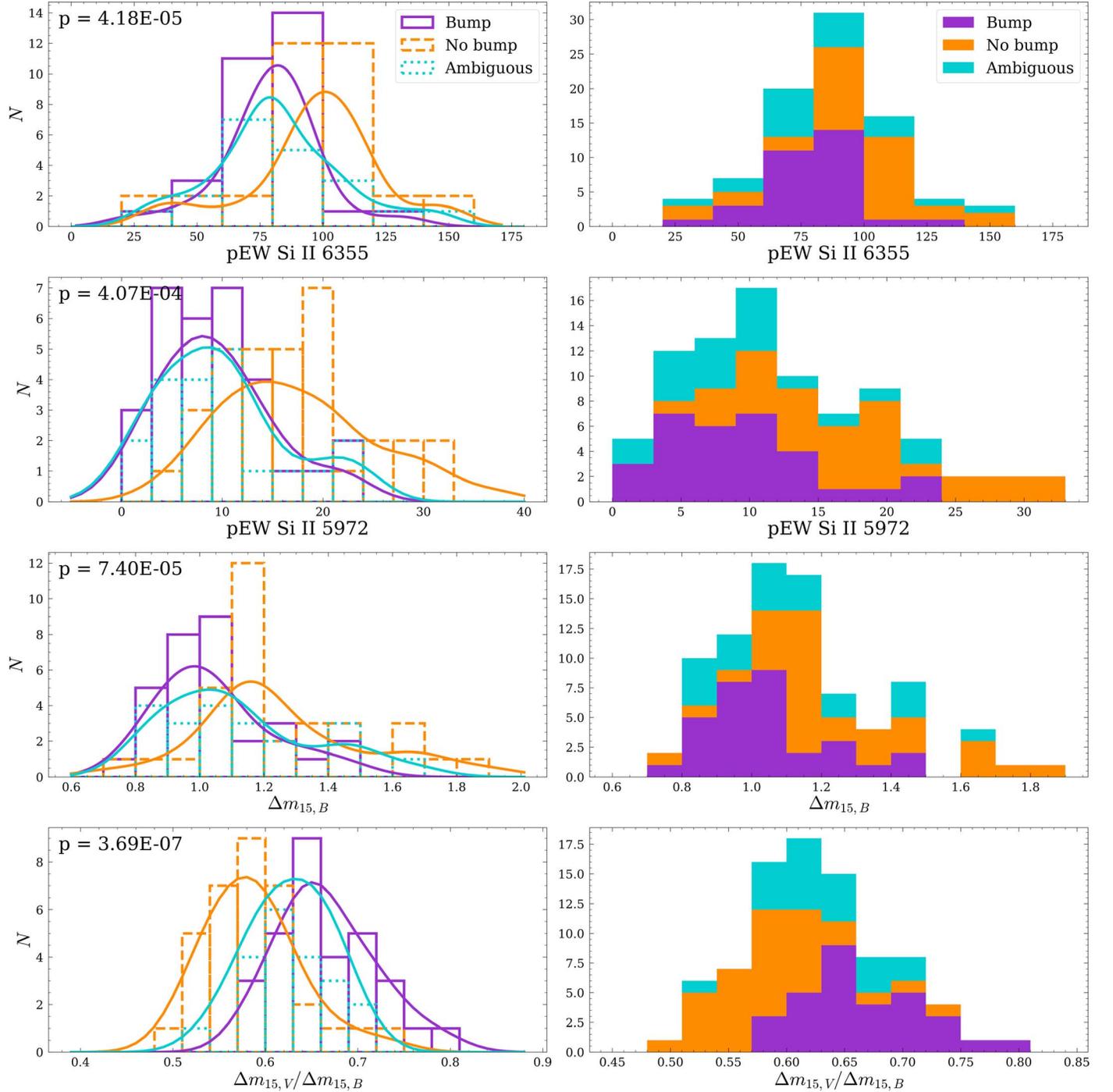

**Figure 3.** Top row: histograms of the pEWs of the Si II λ6355 line. Second row: histograms of the pEW of the Si II λ5972 line. Third row: histograms of the values of the decline rate $\Delta m_{15,B}$. Bottom row: histograms showing the ratios of $\Delta m_{15}$ in the *B* band to $\Delta m_{15}$ in the *V* band. The left column shows overlayed but separate histograms with the sample divided into "bump", "no bump", or "ambiguous" categories. The right column contains stacked histograms to illustrate the combined sample. For all panels, the *p*-values indicate the results of a two-sample KS test run on the "bump" and "no bump" samples. See Table 1 for a complete list of statistical tests and their results. Purple indicates the "bump" sample, yellow indicates the "no bump" sample, and cyan indicates the "ambiguous" sample. The solid lines are kernel density estimates, color-coded the same way as previous, included to aid in the visualization of the separation of the groups.

Although this is a simplified toy model and caution must be taken when interpreting the results, it shows that if inhomogeneous mixing is included, the "bump" feature is reproduced. The occurrence in both *B* and *V* (Figure 6, top left and top right, respectively) support the interpretation of the formation above. Note that inhomogeneous mixing produces a peak brightness that is slightly less than for the unmixed model (Figure 6).

However, this can be compensated for, e.g., by a slightly shorter deflagration phase which would lead to a shift of the theoretical template. Within our framework, the length of the deflagration phase, which is directly related to the shift as well as the brightness, is treated as a free parameter. Thus, we cannot predict the brightness shift within our toy model. In the same realm, the results may apply to other explosion scenarios

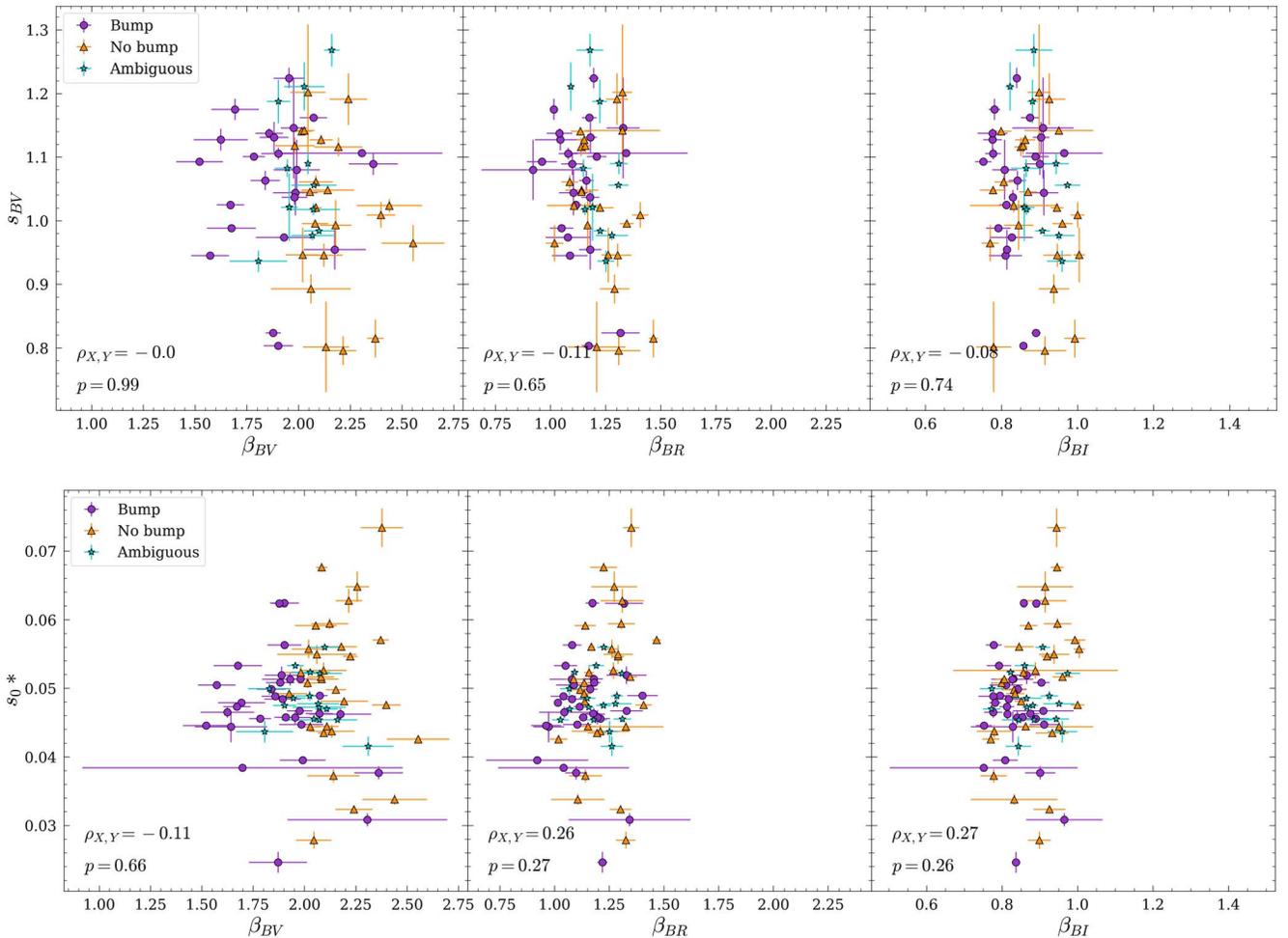

**Figure 4.** Scatter plots showing the lack of correlations between $s_{BV}$ (Burns et al. 2014) and slope $\beta_{XY}$ (top row), and $s_0*$ (Chen et al. 2021) and slope $\beta_{XY}$ (bottom row). $\rho_{X,Y}$ is the Pearson correlation coefficient for each plot, and $p$ is the corresponding $p$-value. We do not find any correlation, indicating that our $\beta_{XY}$ provides different information than $s_{BV}$ and $s_0*$. Further, we do not see any separation between the "bump" and "no bump" groups for neither $s_{BV}$ nor $s_0*$.

for different WD masses if they can produce small-scale chemical inhomogeneities.

The "picket fence" scenario is asymmetric and may lead to intrinsic polarization of the SNe. However, spectropolarimetry of SNe Ia has shown that more luminous objects (small $\Delta m_{15}$) are usually weakly polarized (Wang et al. 2007; Cikota et al. 2019). Recent polarimetry data of SN 1991T-like SNe also show low degrees of polarization (Y. Yang, priv. communication). This can be understood if the size of the "picket fence" projected on the photosphere is much smaller than the size of the photosphere. Large numbers of such picket fences will effectively reduce the observable polarization, as shown by Wang et al. (2007). Based on the models of Wang et al. (2007), the effect of the plumes on the polarization of the emerging light is the strongest when the size of the plume is comparable to that of the photosphere. For the observed upper limit of the polarization of 0.1%–0.2%, the size of the "picket fence" projected onto the photosphere is likely to be around 1/10 of the size of the photosphere, i.e., around 1000 km/sec in velocity space around optical maximum, which is consistent with theoretical expectations (Gamezo et al. 2005; Kromer et al. 2017). The smallness of the plume scale also mitigates the directional dependence of the SN luminosity.

It should also be pointed out that the "picket fence" scenario is particularly appealing for the SNe Ia with tenuous Si layers, such as those found in SN 1991T/SN 1999aa-like objects. In a recent study, Yang et al. (2022) discover that the intrinsic luminosity of SN 1991T/SN 1999aa-like objects is inversely correlated to the pEW of the $\lambda 6355$ Si II Å line at optical maximum. SNe showing bumps in the CMAGIC diagram tend to have weaker Si II lines (Figure 3). The weaker-than-normal $\lambda 6355$ Si II Å feature suggests indeed that the SNe with a "bump" may tend to be more closely related to 91T- or 99aa-like objects than those without a "bump". The Si layer may have been shaped by the instabilities during the deflagration phase, and the effect of the "picket fence" can be more easily detectable for SNe with shallower Si layers.

It is also possible that the "bump" is a result of blueshifting lines due to viewing angle, depending on the velocities of the innermost regions of the ejecta (Maeda et al. 2009). In the "bump" population, our data show that the SN stays bluer for a longer amount of time. If lines from the innermost regions of the ejecta blueshift as a result of their velocities, and the physical origin of the "bump" feature occurs at the Si–Ni interface, it may be a useful early-time indicator of inner ejecta behavior if a SN is viewed from the correct angle.

We emphasize that these calculations are from a 1D model, and 3D models are needed to verify these results. It is possible that a 3D model may reveal asymmetry, i.e., from one viewing angle, a SN shows a "bump" but does not from another. If the

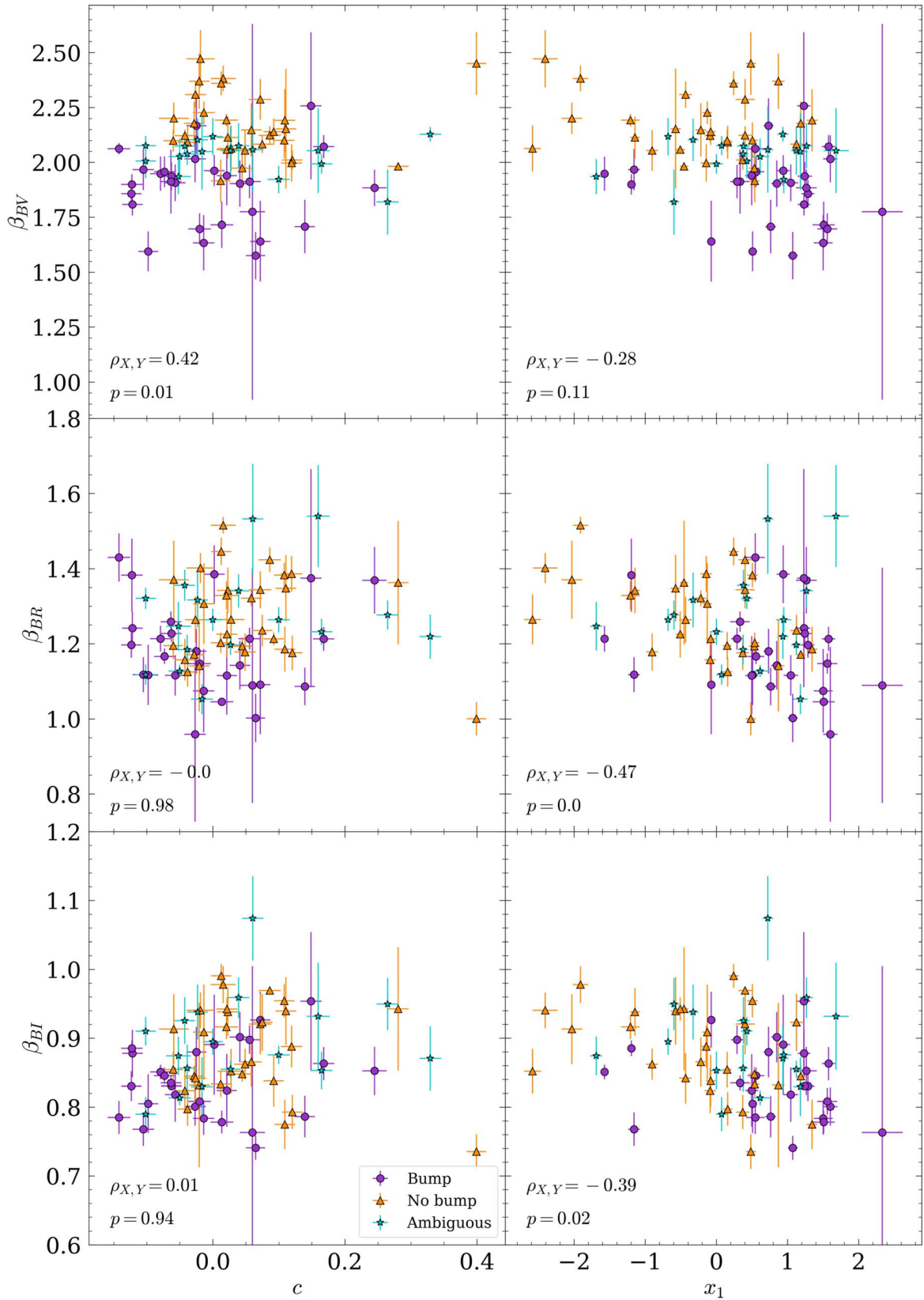

**Figure 5.** Correlations between SALT3 parameters (Kenworthy et al. 2021) and $\beta_{XY}$. $\rho_{X,Y}$ is the Pearson correlation coefficient for all data in each panel, and $p$ is the corresponding $p$-value.

"bump" is related to 3D effects such as asymmetry, it may suggest that the light-curve decline rate is also connected with geometry (Wang et al. 2007; Wang & Wheeler 2008; Maeda et al. 2010; Maund et al. 2010). We also cannot rule out interaction of the ejecta with circumstellar material as the physical cause of the "bump" feature. However, this is unlikely because of the consistent tendency for SNe with a "bump" feature to stay bluer after maximum light, suggesting it is an intrinsic property of a given SN. Because asymmetry introduces intrinsic magnitude and color dispersion among SNe Ia (Wang et al. 2007), it is important to determine the effects, if any, of the "bump" feature on cosmological analyses.

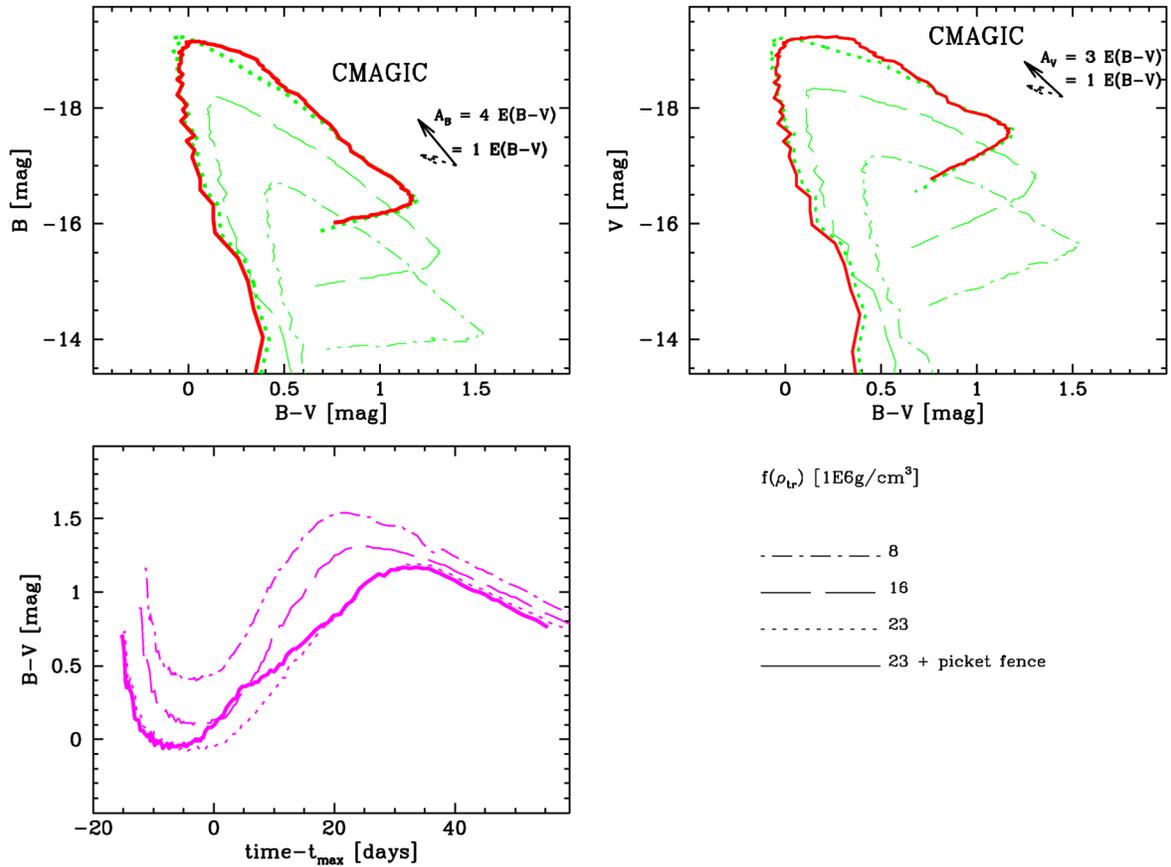

**Figure 6.** Model 23 from Hoeflich et al. (2017), modified to include inhomogeneous mixing. The solid red line is the model with inhomogeneous mixing, while the dotted green line immediately behind the red line is the model without mixing. The other lines are models without mixing at assorted transition densities, $\rho_{tr}$, representing a series from bright to transitional to underluminous SNe Ia with transition densities of 8, 16, and 23 $\times 10^6$ g$^{-1}$ cm$^3$, respectively. CMAGIC templates with different brightness shift along a line defined by the peak brightnesses.

### 3.3. Template Analysis

We manually manipulated three current SN Ia fitting templates in order to check if they accurately reproduced the behavior found in the data. We chose to use fPCA (He et al. 2018), SNooPy (Burns et al. 2014), and SALT3 (Kenworthy et al. 2021).

In order to reproduce the "bump" feature using current templates, we had to keep in mind that the CMAGIC diagram is time-independent if the time axis is stretched or compressed for both light curves. Therefore, we must vary quantities that will produce a change in the CMAGIC diagram, i.e., the shapes of the light curves relative to one another. We chose to vary the width of one band while leaving the width of the other fixed (Figure 3, bottom right) for the fPCA and SNooPy templates.

For the SALT3 template, we did not vary the ratio of the stretch or the time difference between $B_{max}$ and $V_{max}$ because it does not make sense to do so—SALT3 is a spectral template from which synthetic photometry is derived, so the same stretch $x_1$ must apply to all bands, and time is not a free parameter. Thus, we separately varied $x_1$ and $c$ (Figure 9).

Additionally, we fit the SALT3 templates to the data using sncosmo (Barbary et al. 2022), and compared these results to the data- and fPCA-derived quantities (Figure 10).

All templates reproduce the "bump" feature, or lack thereof. However, the fPCA (He et al. 2018) and SNooPy (Burns et al. 2014) templates show a steeper slope in the linear region when a given SN has a "bump", whereas in the data, the SNe with a "bump" tend to have a shallower slope (Figure 2). The SALT3 template (Kenworthy et al. 2021) reproduces the "bump" and the corresponding shallower slope as stretch $x_1$ increases (Figure 9). This implies that as $x_1$ varies, the widths of the light curves in each band do not scale together. Additionally, $\beta_{BV}$ and $\omega_{BV}$ calculated from synthetic SALT3 CMAGIC diagrams show the same patterns as the data- and fPCA-derived quantities (Figure 10), i.e., SNe without a "bump" tend to have a larger slope than those with a "bump".

Going forward, we suggest that future template construction should consider the shape of the CMAGIC diagram while being developed. Otherwise, important indicators of SN physics may be unintentionally excluded.

### 4. Conclusions

The main results of this paper are as follows:

1. The SNfactory data—which do not require k-corrections —confirm the CMAGIC behavior found by Wang et al. (2003).

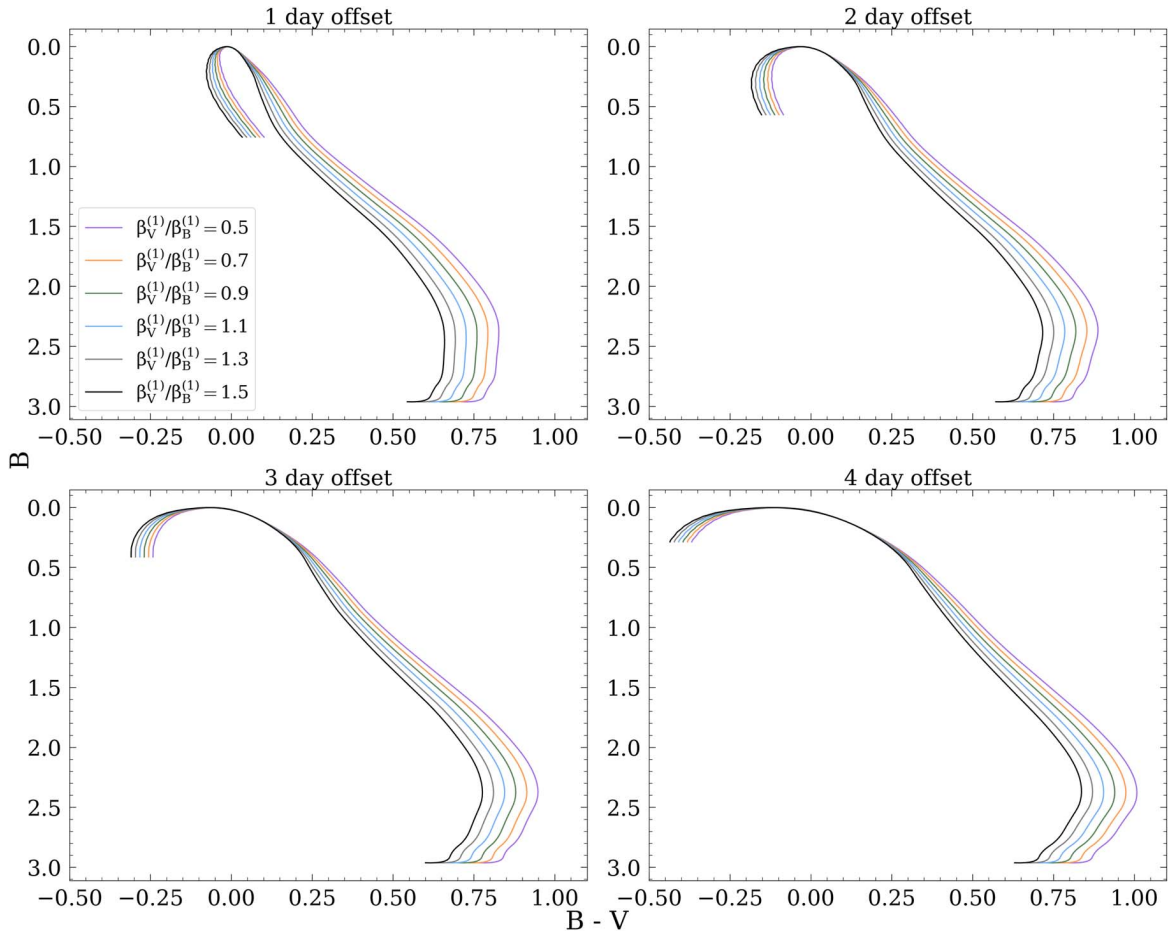

**Figure 7.** CMAGIC diagrams constructed using the fPCA templates from He et al. (2018). Each panel corresponds to a different amount of time between $B_{max}$ and $V_{max}$, such that a 0 day offset is when $B_{max}$ and $V_{max}$ occur at the same time. Consistent with observations, the $V$ band is shifted to later times relative to the $B$ band. The vertical dashed line is located at $(B-V)=0.6$, where $B_{BV}$ is measured. The different colors illustrate the effect of the ratio of the first PC vector, $\beta^{(1)}$, in each band relative to one another. Here, the $B$ band $\beta^{(1)}$ is held fixed at $\beta^{(1)}=1$ and the V-band $\beta^{(1)}$ is allowed to vary. We note that in these templates, a larger offset results in a wider "bump" feature, and a larger ratio of $\beta^{(1)}$ results in a sharper "bump" feature. The $\beta^{(1)}$ parameter in this plot *only* does not represent the slope of the linear region in the CMAGIC diagram; this notation was chosen to be consistent with the notation in He et al. (2018).

2. We defined the "bump" size ($\omega_{XY}$) and found that there is a correlation between the slope ($\beta_{XY}$) and $\omega_{XY}$ in the CMAGIC diagram (Figure 2).
3. We find separation in the spectral and photometric quantities in terms of the presence of a "bump" (Figure 3, Table 1).
4. SNe with a "bump" feature tend to have a slower decline rate than those without a "bump" (Figure 3, third row), confirming the results of Conley et al. (2006). We may circumvent this issue in standardization by using magnitudes chosen based on color rather than time (Wang et al. 2003).
5. The "bump" feature may be caused by mixing of material at the boundary of the Si–Ni region from large-scale Rayleigh–Taylor instabilities (Figure 6), based on 1D DDT models.
6. Current empirical templates are able to reproduce the "bump" feature in the CMAGIC diagram (Figures 7, 8, and 9), and the width of light curves in different photometric bands do not necessarily scale together. The SALT3 template is most reflective of observations because as the stretch, $x_1$, is varied, the slope of the linear region decreases while the "bump" size increases (Figure 9). This is shown with SALT3 fits to our data, as well (Figure 10).

Going forward, we do not recommend thinking of the "bump" feature as a sample-bifurcating property. We believe that the "bump" feature exists on a continuum (see Figures 2, 3, 7, 8, and 9). We base this on the observation that the "ambiguous" category, which includes SNe that may or may not have had a small "bump", always appears between the "bump" and "no bump" categories. From a theoretical perspective, there is no physical reason that there should not be a continuum of Rayleigh–Taylor instability sizes (Section 3.2). We have presented our results in terms of "bump" and "no bump" as an illustrative aid, to explore the point raised by Wang et al. (2003), and to highlight the two extremes of the possible cases.

Future work should include investigating the physical cause of the "bump" feature and its correlation with the slope. For example, it would be interesting to simulate the results of changing the diffusion timescale, amount of mixing, or magnetic field strength. It would also be interesting to combine CMAGIC with polarization studies, to determine observationally if "bump" SNe are more likely to be polarized (Section 3.2). For example, Maund et al. (2010) finds a strong correlation between polarization and the velocity gradient for the Si II λ6355 absorption line. We do not include line velocity gradients in this study; thus, it could be a useful parameter to study in this context if the "bump" feature can be connected with asymmetry and polarization. We also emphasize the need to examine asymmetry with 3D models in relation to the "bump" feature.

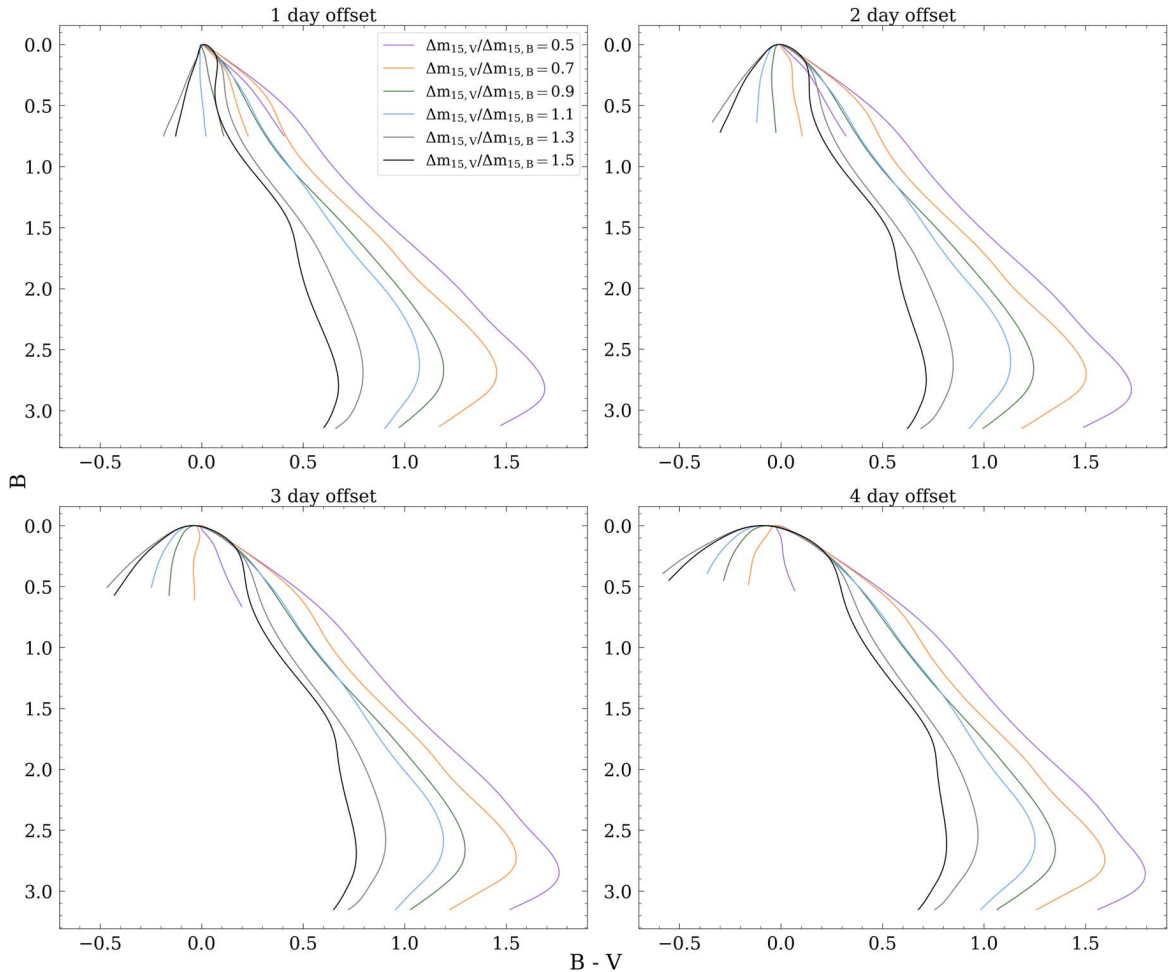

**Figure 8.** CMAGIC diagrams constructed using SNooPy templates (Burns et al. 2011). Each panel corresponds to a different amount of time between $B_{max}$ and $V_{max}$, such that a 0 day offset is means that $B_{max}$ and $V_{max}$ occur at the same time. The vertical dashed line is located at $(B - V) = 0.6$, where $B_{BV}$ is measured. Consistent with observations, the $V$ band is shifted to later times relative to the $B$ band. The different colors illustrate the effect of the ratio of $\Delta m_{15}$ in each band relative to one another. Here, the $B$-band $\Delta m_{15}$ is held fixed at $\Delta m_{15} = 1$ and the $V$-band $\Delta m_{15}$ is allowed to vary. We note that in these templates, a larger offset results in a wider "bump" feature, and a larger ratio of $\Delta m_{15}$ results in a sharper "bump" feature.

These results would be improved by using a data set with more well-sampled SNe Ia in the window 0–60 days after maximum brightness. We may also apply machine-learning techniques, such as the one used by Hu et al. (2022), to fill in gaps in the data.

These results may also have implications for SN Ia cosmology. Wang et al. (2003) shows that the dispersion of standardized magnitudes when using CMAGIC is very small; however, given the results in this work, the dispersion should be reexamined from the perspective of differing CMAGIC diagram morphologies. If it is found that either the "bump" or "no bump" population contributes more to the dispersion than the other, the parameter $\omega_{XY}$ should be incorporated when evaluating scatter. Additionally, because CMAGIC standardizes a magnitude based on measurements over several days rather than the magnitude at a single time, as empirical light-curve models evolve, the effect of their systematic errors propagate through distance modulus calculations. Thus, the CMAGIC diagram may need to be considered when constructing future empirical light-curve templates.

We find that around $\Delta m_{15,B} \approx 1.1$, there is a steep decrease in the number of SNe with a "bump" (Section 3.1). Wang et al. (2006) show that around $\Delta m_{15,B} \gtrsim 1.1$, the Hubble residuals remain consistent with zero. While $\Delta m_{15,B} \lesssim 1.1$, the Hubble residual strays to approximately $-0.6$ mag from zero. Thus, it is necessary to investigate the effect of the "bump" on the Hubble residuals. Based on the results of Wang et al. (2006), we can expect that the "bump" population has a larger Hubble residual dispersion than the "no bump" population.

Conley et al. (2006) notes that it is difficult to observe directly a "bump" for high-$z$ SNe. They find that the systematic effect of the "bump" on cosmological parameters is small, however, because the "bump" is strongly correlated with the slope of the linear region; this issue can now be circumvented without high-cadence observations around maximum brightness. If the "bump" is accounted for within the framework of their analysis, the precision of the results will improve. Conley et al. (2006) determine the effect is small by calculating the probability as a function of stretch that a SN has a "bump" using low-$z$ SNe to determine if this has significant systematic effects at high-$z$. However, it is not currently known if there is a correlation between "bump" incidence and redshift. Additionally, they only needed to determine the systematic effect of the "bump" for four high-$z$ SNe Ia in their sample. While it is true that the effect of the "bump" for these four high-$z$ SNe have a negligible effect on their analysis, this cannot be generalized to all analyses and requires further investigation.

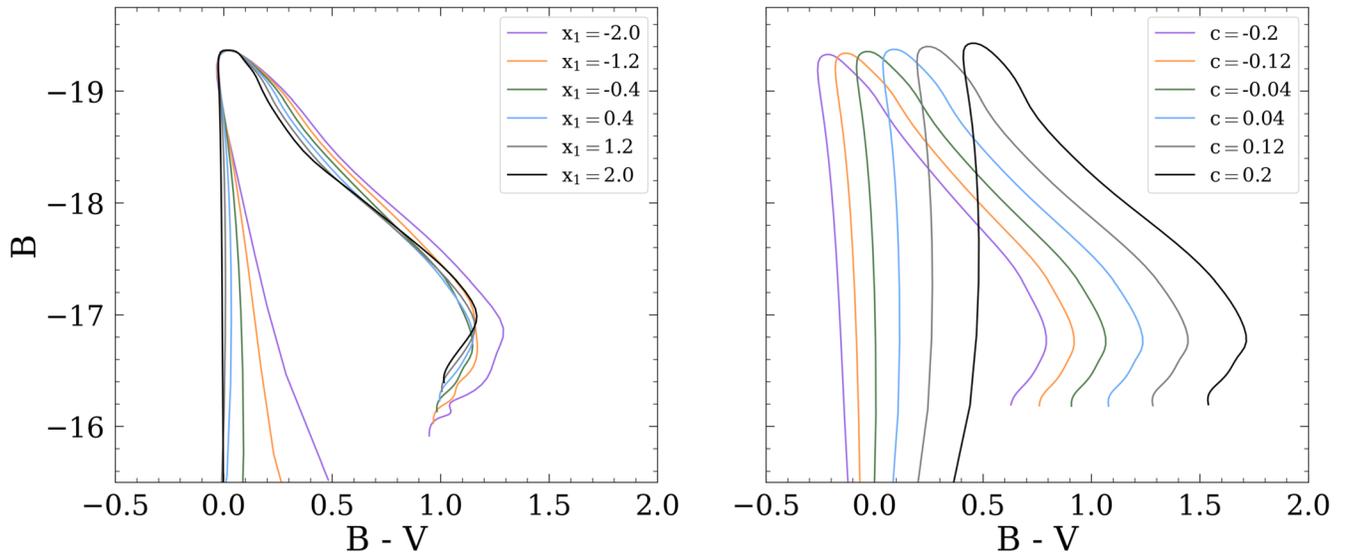

**Figure 9.** CMAGIC diagrams constructed using SALT3 templates (Kenworthy et al. 2021). Left: stretch, $x_1$, is varied. We can see that at larger stretch values, SALT3 reproduces the "bump" feature and the shallower slope that typically appears with it. The color term is held constant with $c = 0$. Right: the color term, $c$, is varied. For smaller values of $c$, the "bump" feature appears. Stretch is held constant at $x_1 = 1$.

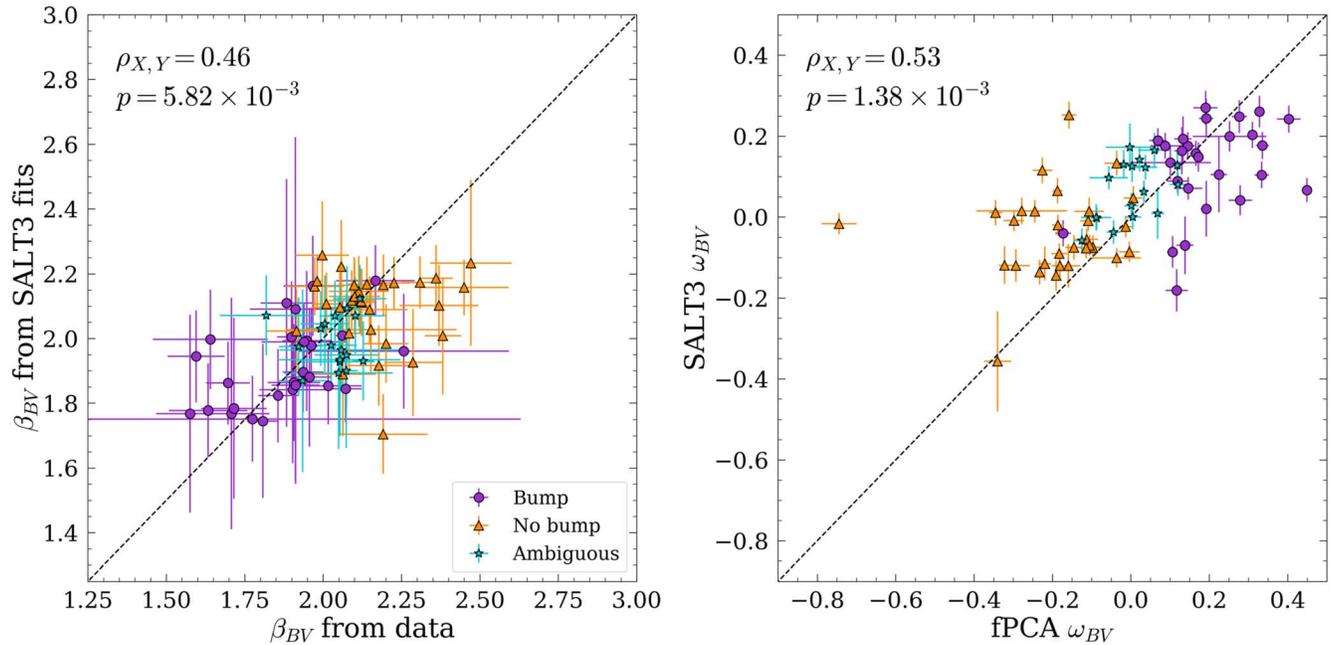

**Figure 10.** Left: slope $\beta_{BV}$ measured from linear fits applied directly to the data compared to $\beta_{BV}$ measured from a linear fit to a synthetic SALT3 CMAGIC diagram in the same phase window. Right: "bump" size $\omega_{BV}$ measured from SALT3 fits and fPCA fits. fPCA $\omega_{BV}$ is calculated using $\beta_{BV}$ from the data rather than a linear fit to the synthetic CMAGIC diagram from the fPCA templates. $\rho_{X,Y}$ is the Pearson correlation coefficient for each plot, and $p$ is the corresponding $p$-value.

L.A. acknowledges support by the NASA Texas Space Grant Consortium and NSF grant AST-1613455. They also thank D'arcy Kenworthy for discussing SALT3, and Yaswant Devarakonda for discussing statistical methods. L.A., L.W., J.Y., and N.S. acknowledge the support of the Mitchell Institute for Fundamental Physics and Astrophysics. P.H. was supported by National Science Foundation (NSF) grant AST-1715133. We also thank the Gordon & Betty Moore Foundation for their support. This work was supported in part by the Director, Office of Science, Office of High Energy Physics of the U.S. Department of Energy under Contract No. DE-AC02-05CH11231. Support in France was provided by CNRS/IN2P3, CNRS/INSU, and PNC and French state funds managed by the National Research Agency within the Investissements d'Avenir program under grant reference numbers ANR-10-LABX-0066, ANR-11-IDEX-0004-02 and ANR-11-IDEX-0007. Additional support comes from the European Research Council (ERC) under the European Union's Horizon 2020 research and innovation program (grant agreement No 759194-USNAC). Support in Germany was provided by DFG through TRR33 "The Dark Universe" and by DLR through grants FKZ 50OR1503 and FKZ 50OR1602. In China support was provided from Tsinghua University 985 grant and NSFC grant No 11173017. Some results were obtained using resources and support from the National Energy Research Scientific Computing Center, supported by the Director, Office of Science, Office of Advanced Scientific Computing Research of the U.S. Department of Energy under Contract No. DE-AC02-05CH11231.

*Facilities:* SuperNova Integral Field Spectrograph (SNIFS, Lantz et al. 2004).

*Software:* astropy (Astropy Collaboration et al. 2018), astrolibpy (Koposov 2017) LCFitter (He et al. 2018; Aldoroty et al. 2022), matplotlib (Hunter 2007) mpfit (Moré 1978; Moré & Wright 1993; Markwardt 2009), numpy (Harris et al. 2020), pandas (McKinney 2010; pandas development team 2020), scipy (Virtanen et al. 2020), sncosmo (Barbary et al. 2022), SNooPy (Burns et al. 2011), uncertainties (Lebigot 2021), HydRa (Hoeflich 2002; Höflich 2009; Hoeflich et al. 2021; Hristov et al. 2021).

## Appendix A
## Synthetic Photometry and Error Propagation

Before generating synthetic photometry, the spectrum is corrected for Milky Way dust and is put in the rest frame using functions from SNooPy (Burns et al. 2014). The area under the spectrum in photon flux units, $F$, in a given filter, is calculated by

$$F = \sum_{i=\lambda_1}^{\lambda_2} \lambda_i f(\lambda_i) R_X(\lambda_i)(\lambda_i - \lambda_{i-1}), \quad (A1)$$

where $f(\lambda_i)$ is the energy flux at a particular wavelength $\lambda_i$ and $R_X$ is the response function for the filter. Then, the variance of $F$ is

$$\sigma_F^2 = \left( \sum_{i=\lambda_1}^{\lambda_2} \lambda_i \sigma_{f(\lambda_i)} R_X(\lambda_i)(\lambda_i - \lambda_{i-1}) \right)^2. \quad (A2)$$

Individual magnitudes are calculated by

$$m = -2.5 \log_{10}\left(\frac{F}{F_{\text{ref}}}\right), \quad (A3)$$

where $F_{\text{ref}}$ is the flux under a given filter for the reference spectrum from Vega (Bohlin et al. 2014). Then, the variance of $m$ is

$$\sigma_m^2 = \frac{\partial m}{\partial f}^2 \sigma_f^2 = \frac{\partial m}{\partial F}^2 \left( \sum_{i=\lambda_1}^{\lambda_2} \frac{\partial F}{\partial f(\lambda_i)} \sigma_{f(\lambda_i)} \right)^2$$

$$= \left(\frac{-2.5}{F \ln 10}\right)^2 \left( \sum_{i=\lambda_1}^{\lambda_2} \lambda_i \sigma_{f(\lambda_i)} R_X(\lambda_i)(\lambda_i - \lambda_{i-1}) \right)^2. \quad (A4)$$

In this work, colors are handled assuming the bands are independent, so errors are calculated using $\sigma_{X-Y} = \sqrt{\sigma_X^2 + \sigma_Y^2}$. This is not the case in reality; however, the results do not change significantly if the covariance terms are included. If covariance were considered, the errors would be calculated using $\sigma^2 = JCJ^T$, and would be done as follows.

For arbitrary color $X - Y$

$$m_X - m_Y = -2.5 \log\left(\frac{F_X}{F_{\text{ref},X}}\right) + 2.5 \log\left(\frac{F_Y}{F_{\text{ref},Y}}\right). \quad (A5)$$

For ease of calculations, we will treat $m_X - m_Y$ as a function of $f(\lambda_i)$. Then, for arbitrary color $X - Y$, the Jacobian is

$$\boldsymbol{J}_{X-Y} = \left[ \frac{\partial m_{X-Y}}{\partial f(\lambda_0)} \ \frac{\partial m_{X-Y}}{\partial f(\lambda_1)} \ \cdots \ \frac{\partial m_{X-Y}}{\partial f(\lambda_N)} \right], \quad (A6)$$

where $N$ is the last measured wavelength in the spectrum. The $i$th entry is

$$\frac{\partial m_{X-Y}}{\partial f(\lambda_i)} = -\frac{1.09}{F_X} R_X(\lambda_i)(\lambda_i - \lambda_{i-1}) + \frac{1.09}{F_Y} R_Y(\lambda_i)(\lambda_i - \lambda_{i-1}). \quad (A7)$$

Then, the covariance matrix is diagonal, and each entry is the spectrum error provided in the data

$$\boldsymbol{C} = \begin{bmatrix} \sigma_{f(\lambda_0)}^2 & & & \\ & \sigma_{f(\lambda_1)}^2 & & \\ & & \ddots & \\ & & & \sigma_{f(\lambda_N)}^2 \end{bmatrix}. \quad (A8)$$

## Appendix B
## fPCA Fitting and Error Propagation

Light curves are described by

$$g_{s\lambda}(q) = m_{s\lambda} + \phi_{0\lambda}(q) + \sum_{k=1}^{K} \beta_{s\lambda}^{(k)} \phi_{k\lambda}(q), \quad (B1)$$

where $q$ is the phase, $m_{s\lambda}$ is the peak magnitude, $s$ is the corresponding SN, $\lambda$ is the corresponding filter, $\phi_{k\lambda}$ are the fPCA vectors, and $\beta_{s\lambda}^{(k)}$ are the outputs of the fPCA analysis code (He et al. 2018). The errors in the entire light curve can be calculated with $\sigma^2 = JCJ^T$. The covariance matrix is output from the fPCA analysis code.

The Jacobian is

$$\boldsymbol{J}_{g_{s\lambda}} = \left[ \frac{\partial g_{s\lambda}}{\partial q_0} \ \frac{\partial g_{s\lambda}}{\partial m_{s\lambda}} \ \frac{\partial g_{s\lambda}}{\partial \beta_{s\lambda}^{(1)}} \ \frac{\partial g_{s\lambda}}{\partial \beta_{s\lambda}^{(2)}} \ \frac{\partial g_{s\lambda}}{\partial \beta_{s\lambda}^{(3)}} \ \frac{\partial g_{s\lambda}}{\partial \beta_{s\lambda}^{(4)}} \right],$$

or, equivalently

$$\boldsymbol{J}_{g_{s\lambda}} = \begin{bmatrix} 0 & 1 & \phi_{1\lambda}(q) & \phi_{2\lambda}(q) & \phi_{3\lambda}(q) & \phi_{4\lambda}(q) \end{bmatrix}.$$

The error in $\Delta m_{15}$ is calculated similarly. If

$$\Delta m_{15} = g_{s\lambda}(q_0 + 15) - g_{s\lambda}(q_0) = \phi_{0\lambda}(q_0 + 15) - \phi_{0\lambda}(q_0)$$
$$+ \sum_{k=1}^{K} \beta_{s\lambda}^{(k)} (\phi_{k\lambda}(q_0 + 15) - \phi_{k\lambda}(q_0)), \quad (B2)$$

then $\boldsymbol{J}_{\Delta m_{15}}$ is given by

$$\boldsymbol{J}_{\Delta m_{15}} = \left[ \frac{\partial \Delta m_{15}}{\partial q_0} \ \frac{\partial \Delta m_{15}}{\partial m_{s\lambda}} \ \frac{\partial \Delta m_{15}}{\partial \beta_{s\lambda}^{(1)}} \ \frac{\partial \Delta m_{15}}{\partial \beta_{s\lambda}^{(2)}} \ \frac{\partial \Delta m_{15}}{\partial \beta_{s\lambda}^{(3)}} \ \frac{\partial \Delta m_{15}}{\partial \beta_{s\lambda}^{(4)}} \right].$$

The first entry is

$$\frac{\partial \Delta m_{15}}{\partial q_0} = \frac{\partial \phi_0(q_0 + 15)}{\partial q_0} - \frac{\partial \phi_0(q_0)}{\partial q_0}$$
$$+ \sum_{k=1}^{K} \beta_{s\lambda}^{(k)} \left( \frac{\partial \phi_{k\lambda}(q_0 + 15)}{\partial q_0} - \frac{\partial \phi_{k\lambda}(q_0)}{\partial q_0} \right). \quad \text{(B3)}$$

These values can be calculated directly when this formula is discretized because each vector exists in the form of a discrete grid. The second entry is always 0. The derivatives with respect to $\beta_{s\lambda}^{(k)}$ are

$$\frac{\partial \Delta m_{15}}{\partial \beta_{s\lambda}^{(k)}} = (\phi_{k\lambda}(q_0 + 15) - \phi_{k\lambda}(q_0)). \quad \text{(B4)}$$

The covariance matrix used is the same one that is output from the fPCA analysis code. The fPCA templates and a fitting routine can be used by installing snlcpy, a Python package located at https://github.com/laldoroty/snlcpy (Aldoroty et al. 2022).

## Appendix C
## Bump Size Error Propagation

The "bump" size is defined as

$$\omega = (\beta_{XY}(X - Y)_{\max} + X_{XY0}) - m_{X,\max}, \quad \text{(C1)}$$

where $m_{X,\max}$ is the magnitude at maximum brightness and $(X - Y)_{\max}$ is the color that corresponds to this time.

Again, we use $\sigma^2 = JCJ^T$ to calculate the error in "bump" size. We construct a covariance matrix with rows and columns for $m_X$, $X - Y$, $\beta_{XY}$, and $X_{XY0}$ as

$$C_{\text{bump}} = \begin{bmatrix} \sigma^2_{m_{X,\max}} & \sigma^2_{m_{X,\max}(X-Y)_{\max}} & 0 & 0 \\ \sigma^2_{m_{X,\max}(X-Y)_{\max}} & \sigma^2_{(X-Y)_{\max}} & 0 & 0 \\ 0 & 0 & \sigma^2_{\beta_{XY}} & \sigma^2_{\beta_{XY}X_{XY0}} \\ 0 & 0 & \sigma^2_{\beta_{XY}X_{XY0}} & \sigma^2_{X_{XY0}} \end{bmatrix}. \quad \text{(C2)}$$

If it is assumed that the X and Y bands are independent, then $\sigma^2_{m_{X,\max}(X-Y)_{\max}} = 0$. Although the bandpasses are not independent in reality, we assume independence in this work for ease of calculation, as the final results would not be significantly affected by this complication. The Jacobian is

$$J_\omega = \begin{bmatrix} \frac{\partial \omega}{\partial m_{X,\max}} & \frac{\partial \omega}{\partial (X-Y)_{\max}} & \frac{\partial \omega}{\partial \beta_{XY}} & \frac{\partial \omega}{\partial X_{XY0}} \end{bmatrix} \quad \text{(C3)}$$

$$= \begin{bmatrix} -1 & \beta_{XY} & (X-Y)_{\max} & 1 \end{bmatrix}. \quad \text{(C4)}$$

## ORCID iDs

L. Aldoroty ⓘ https://orcid.org/0000-0003-0183-451X
L. Wang ⓘ https://orcid.org/0000-0001-7092-9374
P. Hoeflich ⓘ https://orcid.org/0000-0002-4338-6586
J. Yang ⓘ https://orcid.org/0000-0002-1376-0987
N. Suntzeff ⓘ https://orcid.org/0000-0002-8102-181X
P. Antilogus ⓘ https://orcid.org/0000-0002-0389-5706
C. Aragon ⓘ https://orcid.org/0000-0002-9502-0965
S. Bailey ⓘ https://orcid.org/0000-0003-4162-6619
C. Baltay ⓘ https://orcid.org/0000-0003-0424-8719
K. Boone ⓘ https://orcid.org/0000-0002-5828-6211
C. Buton ⓘ https://orcid.org/0000-0002-3780-7516
Y. Copin ⓘ https://orcid.org/0000-0002-5317-7518
S. Dixon ⓘ https://orcid.org/0000-0003-1861-0870
D. Fouchez ⓘ https://orcid.org/0000-0002-7496-3796
E. Gangler ⓘ https://orcid.org/0000-0001-6728-1423
R. Gupta ⓘ https://orcid.org/0000-0003-1820-4696
B. Hayden ⓘ https://orcid.org/0000-0001-9200-8699
Mitchell Karmen ⓘ https://orcid.org/0000-0003-2495-8670
A. G. Kim ⓘ https://orcid.org/0000-0001-6315-8743
M. Kowalski ⓘ https://orcid.org/0000-0001-8594-8666
D. Küsters ⓘ https://orcid.org/0000-0002-9207-4749
P.-F. Léget ⓘ https://orcid.org/0000-0002-8357-3984
J. Nordin ⓘ https://orcid.org/0000-0001-8342-6274
R. Pain ⓘ https://orcid.org/0000-0003-4016-6067
S. Perlmutter ⓘ https://orcid.org/0000-0002-4436-4661
K. A. Ponder ⓘ https://orcid.org/0000-0002-8207-3304
D. Rabinowitz ⓘ https://orcid.org/0000-0003-4961-2653
M. Rigault ⓘ https://orcid.org/0000-0002-8121-2560
D. Rubin ⓘ https://orcid.org/0000-0001-5402-4647
C. Saunders ⓘ https://orcid.org/0000-0002-4094-2102
G. Smadja ⓘ https://orcid.org/0000-0002-9093-8849
N. Suzuki ⓘ https://orcid.org/0000-0001-7266-930X
R. C. Thomas ⓘ https://orcid.org/0000-0002-2834-4257